\documentclass[twocolumn,preprintnumbers,amsmath,amssymb]{revtex4}
\usepackage{graphicx}
\usepackage{longtable}

\begin{document}

\title{A new approach to two-level model calculation of isospin mixing in nuclei }

\author {Sukhendusekhar Sarkar 
}
\affiliation{Department of Physics, Indian   Institute   of   Engineering  Science  and
Technology, Shibpur, Howrah - 711103, INDIA} 
\thanks{ss@physics.iiests.ac.in}

\date{\today}

\begin{abstract}
 A new method has been proposed for isolated two-level model to calculate isospin mixing probability in nuclei overcoming common limitations of usual shell model results with isoscalar nuclear Hamiltonian. The method is based on locating the unperturbed levels of the mixed doublet before mixing. Experimental and shell model level energies, electromagnetic/Gamow-Teller transition matrix elements associated with a doublet or two doublet pairs in a nucleus are used to calculate isospin mixing probability in seven self-conjugate nuclei. The four self-conjugate nuclei (P, S, Cl, Ar) considered here show large isospin mixing matrix elements and large unperturbed energy-gaps of the observed isospin-mixed doublet pairs. Large isospin mixing (31.40-48.76 \%) is found in the observed doublet ($1^{+},1^{+}$) at 9828.11 and 9967.19 keV, respectively, in $ ^{24}Mg$. This is probably the largest isospin mixing ever found in a nucleus. This is much larger than  that
 $({20}_{-9.904}^{+9.142}$\%) has been found recently in $^{26}$Si. The method is general enough to be applicable to other two-level/multi-level mixing problems and in particular, might be useful for consideration of isospin mixing in the context of Fermi beta decay also.

\end{abstract}

\maketitle

Understanding the structure of some quantum systems in an isolated two-level model is widely popular and is quite fruitful.
Consideration of a quantum two-level model perhaps goes back to 1916 when Einstein derived the Planck formula for the distribution of energy 
density u($\nu$,T) in the spectrum of a black-body in a two-level model and introduced the concept of spontaneous emission.
Many other applications of two-level system, like Stern-Gerlach system, Rabi model, Lipkin 
 model, to name a few, are found subsequently. The states of a quantum system are labeled by a set of quantum numbers corresponding to a complete set of commuting operators, representing observables associated with the system. This set of quantum numbers are postulated to furnish complete information about the system. Each of these commuting observables represents a symmetry associated with the system and the corresponding quantum number is used as a label for the states of the system. For dynamical reasons one ( or more) of these symmetries can be broken and the corresponding quantum  number is no longer a good quantum number. Concept of isospin quantum number was introduced by W. Heisenberg \cite{heisenberg}
 to distinguish between proton and neutron, as the two different charge states of the same particle 'nucleon'(N). From the mirror symmetry and the isobaric multiplet symmetry in the spectra of nuclei, the charge symmetry and charge independence, respectively, of the strong nucleon-nucleon interaction potential V(NN) were established. Thus introducing nuclear isospin quantum numbers T and its z-component T$_z$, representing identity of the nucleus, corresponding to charge independence and charge symmetries, respectively, one labels a nuclear state, for example, by $|E,J^\pi $,T,T$_z>$, where J$^\pi$ is the total angular momentum and parity of the state and E is the energy of the state.

A very popular application of two-level mixing is in the context of nuclear shell model (SM) \cite{Bru}, \cite{Casten} configuration mixing and isospin mixing in nuclear states due to isospin symmetry breaking by the perturbation ($V_{CD}$) \cite{ekw}, the charge-dependent coulomb and nuclear potentials. The mixing leads to isospin-forbidden/retarded electromagnetic (EM) transition \cite{ekw} to occur between levels of two pairs of isospin mixed doublets in a nucleus and this has been used to measure the isospin mixing probability (b$^2$).\\

 In this Letter a new approach to the two-level mixing model has been presented which is capable of calculating  isospin mixing probability in an easy semi-empirical way, circumventing hurdles of using isoscalar Hamiltonian for isospin mixing in nuclear shell model.
The method uses experimental level energies, E$_1$ and E$_2$ of the pairs of mixed doublets, measured reduced transition probabilities, forbidden B(E1) in particular, and the corresponding quantities from SM calculations. In some cases, only one pair of mixed doublet is of concern in a nucleus.  
 The new method introduced in this work is based on locating the unperturbed energy levels, that is, locations of the observed isospin-mixed doublet before mixing  and also the precise expression for the mixing matrix element. It is found that the isospin mixing probability, is proportional to the gap between the unperturbed energy levels.

 If the isospin symmetry is broken, the observed mixed doublet states $|E_1, J^\pi>$ and $|E_2,J^\pi>$ of the nucleus are then a linear superposition of the orthonormal unperturbed basis states $|H_{11},J^\pi,T=T-1>$ and $|H_{22},J^\pi, T=T>$, having pure T=T-1 and T=T, respectively. $H_{11}$, $H_{22}$ are the unperturbed level energies corresponding to the experimental energies $E_1$, $E_2$ of the observed doublet. $H_{11}$ = $<H_{11},J^\pi,T=T-1|H|H_{11},J^\pi,T=T-1>$, $H_{22}$ = $<H_{11},J^\pi,T=T|H|H_{11},J^\pi,T=T>$, H is the Hamiltonian of the two-level system.

We shall henceforth consider only the mixing of pure T = 0 and T = 1 states. One can thus write for the wave functions of the observed doublet in terms of unperturbed wave functions $|H_{11},J^\pi,T=0>$ and $|H_{22},J^\pi, T=1>$, abbreviated to $|J^\pi,T=0>$ and $|J^\pi, T=1>$, as 
\begin{equation}
%$$
	\begin{bmatrix} 
	|E_1,J^\pi>  \\
	|E_2,J^\pi> \\
	\end{bmatrix}
=
     \begin{bmatrix}
         \sqrt{1-b^2} & -|b| \\
         |b| & \sqrt{1-b^2} \\ 
             \end{bmatrix}
 \times     
     \begin{bmatrix}
         |J^\pi, T=0> \\
         |J^\pi, T=1> \\ 
             \end{bmatrix}
	\quad
%	$$
\end{equation}
Here $|b|$ is the isospin mixing amplitude. We have assumed $|E_1$,$J^\pi>$) as the lower eigenvalue and $|b|$ as the smaller amplitude. Thus $|E_1,J^\pi>$ is predominantly T =0 and $|E_2,J^\pi>$ is predominantly T = 1.  
Changing the basis to express the unperturbed states 
in the basis set formed by the observed doublet states, one can write,
\begin{equation}
        \begin{bmatrix}
         |J^\pi, T=0>   \\
         |J^\pi, T=1>   \\
          \end{bmatrix}
=
        \begin{bmatrix}
            \sqrt{1-b^2}   & |b|   \\
            -|b|  &  \sqrt{1-b^2}   \\
           \end{bmatrix}
    \times
          \begin{bmatrix}
           |E_1,J^\pi>   \\
           |E_2,J^\pi>   \\
           \end{bmatrix}
       \quad
\end{equation}
Shell model (SM) calculation with an isoscalar Hamiltonian $H^{(0)}$ can give eigenenergies $E^{(0)}_{1}$ and $E^{(0)}_{2}$ corresponding to the experimental energies $E_1$ and $E_2$ of the isospin mixed doublet. However, such theoretical results for the energy eigenvalues may not have the properties,

\begin{equation}
(i)\ Tr^{SM} = {E_{1}}^{(0)} + {E_{2}}^{(0)} =   E_1 + E_2 = Tr^{ex} 
\end{equation}
Here Tr = Trace.

\begin{equation}
(ii) \   \Delta^\prime  = (E_2 - E_1) \geq \Delta = (E^{(0)}_{2} -E^{(0)}_{1})
\end{equation}

Many SM model calculations performed  in the present work in the sd and fp shells (see also \cite{Lise}) show this situation (Table I \& text). In fact, the doublet states for which isospin mixing is present, their energy eigenvalues are not in principle obtainable from SM calculation with isoscalar Hamiltonian because of the charge-dependent perturbation $V_{CD}$. Experimental level energies of the doublet contain also the non-isoscalar contribution from $V_{CD}$ which is reflected in the  trace difference ((Eq(16), below).

One may try to improve upon  $E_{1}^{(0)}$ and $E_{2}^{(0)}$ by diagonalising the $2\times 2$ matrix, 

\begin{equation}
       H \\
=
    \begin{bmatrix}
E_{1}^{(0)}  & H_{12}   \\
H_{12}  & E_{2}^{(0)}   \\
\end{bmatrix}
\end{equation}

to get shifted eigenvalues, ${\lambda}_{\pm}$ = 1/2[${Tr}^{SM}$ $\pm$ $\sqrt{\Delta^2 + {(2H_{12})}^2}$] where,
 $H_{12}$  is the mixing matrix element. Even if ${H}_{12}$ is  obtained experimentally, 
 even then, inequality of ($\lambda_+$ + $\lambda_-$) and (${Tr}^{ex}$) may remain and the gap between the shifted eigenvalues $\sqrt{\Delta^2 + (2H_{12})^2}$ may not be equal to $\Delta^{\prime}$.

Using the unperturbed
basis set of Eq(2) one can construct the $2 \times 2$ matrix for H, with $<E_1,J^\pi|H|E_1,J^\pi> = E_1$ and $<E_2,J^\pi| H|E_2,J^\pi> = E_2$,

\begin{equation}
       H \\
=
    \begin{bmatrix}
E_1 + b^2(E_2-E_1)  & \sqrt{b^2-b^4} (E_2-E_1)   \\
\sqrt{b^2-b^4} (E_2-E_1)  & E_2 - b^2(E_2-E_1)   \\
\end{bmatrix}
\end{equation}

The diagonalisation of this matrix obviously gives the experimental level energies.
From the matrix Eq(6), one can identify $H_{11} = E_1 + b^2(E_2 - E_1)$ and $H_{22} = E_2 - b^2 (E_2 - E_1)$ and they are the exact locations of the levels before mixing.
 One can see clearly from these expressions for the unperturbed level energies, that $H_{11}$ and  $H_{22}$ were up and down, respectively, by the amount $b^2 \Delta^{\prime}$, with respect to their respective observed energies $E_1$ and $E_2$. 

That $H_{11}$ and $H_{22}$ are actual unperturbed energies, is clearly demonstrated by Eq(6), namely, $H_{11}$ + $H_{22}$ = $E_1$ + $E_2$, the trace invariance. 
Locating the unperturbed energy levels allows one to get the gap as, 
\begin{equation}
\bar{\Delta}  = H_{22} - H_{11} = (1-2b^2) \Delta^\prime
\end{equation}
 showing that $\bar{\Delta} \leq \Delta^\prime $ always and the most important relation between $b^2$ and $\bar{\Delta}$ 
\begin{equation}
b^2 = (1-{\bar{\Delta}}/\Delta^\prime)/2
\end{equation}
Also,
\begin{equation}
b^2 = (H_{11}-E_1)/\Delta^\prime = (E_2-H_{22})/\Delta^\prime
\end{equation}

From the expressions of $H_{11}$ and $H_{22}$ one can derive a new expression for the isospin mixing probability ($b^2$),
\begin{equation}
(b^2 - b^4) = |{(H_{11}H_{22} - E_1E_2)}|/{\Delta^\prime}^2
\end{equation}

 Eq(6) gives the off-diagonal isospin mixing matrix element as,
\begin{equation}
H_{12} = \sqrt{(b^2  - b ^4 )}(E_2  - E_1) = \sqrt{(b ^2  - b ^4 ) } \Delta^\prime
\end {equation}\\

One can also get $H_{12}$ = $\sqrt{|H_{11}H_{22}- E_{1}E_{2}|}$ from Eq(10).

When the unperturbed levels are degenerate,
 $|b| = \sqrt{|(1-b^2)|} = 1/\sqrt{2}$, equal mixing amplitude or probability for each level. The symmetric energy shift is,
\begin{equation}
 \Delta E_{S} = (\Delta^{\prime}- \bar{\Delta)}/2
\end{equation}
The level repulsion P is given by,
\begin{equation}
P = (H_{11} - E_{1})/(H_{22}-H_{11})
\end{equation}

\begin{equation}
P= b^2/(1-2b^2)
\end{equation}

P becomes maximum at b = $\pm$ (1/2) with a value 1/2.\\ 

Crucial point is that when $b^2$ is not known experimentally but $E_1$ and $E_2$ are, one can still locate the unperturbed levels of a doublet using results of SM calculation of energies and EM/Gamow-Teller (GT) transition matrix elements. In order to circumvent the difficulties with the usual SM results as mentioned in Eq(3)-Eq(4), we define,
\begin{equation}
\bar{\delta} = \Delta + n |{\Delta}Tr|
\end{equation}
where, 
\begin{equation} 
{\Delta}Tr = {Tr}^{ex} - {Tr}^{SM},
\end{equation}
Here n is any number, positive, negative or zero that brings $\bar{\delta}$ below $\Delta^{\prime}$ and satisfy certain criteria, depending on the SM result,
to be derived below. It is to be noted that $\Delta$ obtained from SM eigenvalues (Eq(4)) for the doublet can be $\geq$ or $\leq$ $\Delta^{\prime}$. To locate the unperturbed levels, we also define,
\begin{equation}
\bar{E_1} = (1/2)(Tr^{ex}-\bar{\delta}); 
  \bar{E_2} = (1/2(Tr^{ex}+ \bar{\delta})
\end{equation}
Obviously, the energies $\bar{E_1}$, $\bar{E_2}$, satisfy trace-invariance  and ($\bar{E_2}$ - $\bar{E_1}$) = $\bar{\delta}$ can be made always $\leq$ ${\Delta}^{\prime}$ by
proper choice of n in Eq(15). Symmetric shift is also ensured from the formula Eq(12) with $\bar{\Delta}$ replaced by $\bar{\delta}$. Thus, $\bar{\delta}$ and ($\bar{E_1}$ , $\bar{E_2}$), though variable at this stage, giving variable $b^2$, are playing the roles of $\bar{\Delta}$ and ($H_{11}$ , $H_{22}$), respectively. Their actual values can be fixed once $b^2$ is obtained. Using Eq(10) with a replacement of $H_{11}$$H_{22}$ by $\bar{E_1}$$\bar{E_2}$, one can show, with $H_{12}$ 
given by Eq(11), and $\bar{E_1}$ + $\bar{E_2}$ = $E_1$ + $E_2$, that the diagonalisation of the matrix,

\begin{equation}
       H \\
=
    \begin{bmatrix}
\bar{E_{1}}  & H_{12}   \\
H_{12}  & \bar{E_{2}}   \\
\end{bmatrix}
%\quad
\end{equation}
always gives eigenvalues in the form, $E_2$,$E_1$ = (1/2)($Tr^{ex}\pm \Delta^{\prime}$), for the allowed range of values of n to be calculated as prescribed below. Thus the problem of using energy eigenvalues from SM calculation with an isoscalar Hamiltonian is solved by using semi-empirical input ${\Delta}$Tr = ${Tr}^{ex}$ - ${Tr}^{SM}$.  
Choosing a value of n within its range, $\bar{\delta}$ can be calculated from Eq(15) and $b^2$ can be obtained using Eq(8) by replacing $\bar{\Delta}$  by $\bar{\delta}$, that is, by expressing  $b^2$ as,
\begin{equation}
b^2 = (1/2)[ (1-\Delta/\Delta^{\prime})- n|\Delta{Tr}|/\Delta^{\prime}]
\end{equation} 
Here, n $\leq$ 0 if $\Delta\geq\Delta^{\prime}$ and n$\geq$0 if $\Delta\leq\Delta^{\prime}$. The range, $n_c$ $\leq $n$ \leq$ $n_L$ can be obtained from the conditions that $b^2 \geq 0$ and $b^2 \leq  1/2$, for each doublet for a particular SM calculation.

Isospin mixing probability $b^2$ has been calculated using the method prescribed above for seven self-conjugate nuclei, ($^{30}P$,$^{32}S$,$^{34}Cl$,$^{36}Ar$) and ($^{24}Mg$,$^{54}Co$,$^{64}Ge$). For the first group, two pairs of isospin-mixed doublets, that is, [$E_{k}$(${J_{i}}^{{\pi}_{1}}$)] and [$E_{k}$(${J_{f}}^{{\pi}_{2}}$)], (k = 1,2) in each nucleus are to be considered. Experimental level energies ($E_1,E_2$) for all those pairs in each of the first three nuclei are known. For $^{36}Ar$, upper level energy $E_2$(${4_1}^{+}$, T = 1) of the final (lower) pair is not yet  known. Equating ${Tr}^{SM}$ = ${Tr}^{ex}$, level energy is estimated to be at 10460.6 keV. Specifically, isospin forbidden E1 transitions used here are, ${4_1}^{-}\rightarrow {3_1}^{+}$  2259 keV transition in $ ^{30}$P,  ${3_1}^{-}\rightarrow {2_1}^{+}$  2776 keV transition in $ ^{32}$S,  ${4_1}^{-}\rightarrow {3_1}^{+}$  3454 keV transition in $ ^{34}$Cl and  ${5_1}^{-}\rightarrow {4_1}^{+}$  757 keV transition in $ ^{36}$Ar (Table I). 
 
The transitions are always from the lower $|E_{1}$,${J_{i}}^{{\pi}_{1}}>$ level of the upper pair to lower $|{E_1},{J_f}^{{\pi}_{2}}>$ level of the lower pair in each of the four nuclei. ${b_i}^2$ and ${b_f}^2$ depend on $n_i$ and $n_f$ through Eq(19) are to be obtained by fitting $M(E1)_{theory} $(RHS of Eq(20)) to the experimental transition matrix element ${M(E1)}_{expt}$ obtained from ${B(E1)}_{expt}$ ($e^2 {fm}^2$), $M(E1) = \sqrt{(2J_i+1)B(E1)}$. We set the limit $|{M(E1)}_{expt}-{M(E1)}_{theory}|$$\le$ $10^{-3}$ or ($|{M(E1)}_{expt}|- |{M(E1)}_{theory}|$)$\le$ $10^{-3}$.

 Since two parameters (${b_i}^2$ and ${b_f}^2$) are involved and only one measured ${M(E1)}_{expt}$ to fit, one can calculate (${b_i}^2$ , ${b_f}^2$) pair for two doublets by varying $n_i$ and $n_f$, with small increments in their respective ranges, starting from the  minimum values of ${b_i}^2$ and ${b_f}^2$
 and comparing $M(E1)_{theory}$ with the experimental result each time. After a few  repetitions of the process, one achieves the desired fit. Then  a finer tuning gives the minimum (${b_i}^2$, ${b_f}^2$) set consistent with minimum $H_{12}$ for each doublet and also consistent with certain limiting values given below.
 
$M(E1)_{theory}$ calculated from SM uses the expression 
(RHS (of Eq(20)), for the first group,

\begin{equation}
M(E1)_{expt} = - \sqrt{(1-{b_f}^2)} |b_i| m_2 - \sqrt{(1-{b_i}^2} |b_f| m_1
\end{equation}
where, the reduced transition matrix elements
$m_2 =  <{J_f}^{{\pi}_f},T=0|E1|{J_i}^{{\pi}_i}, T=1>$ and 
$m_1 = <{J_f}^{{\pi}_f},T=1|E1|{J_i}^{{\pi}_i}, T=0>$.
$m_1$ and $m_2$ are obtained from SM calculations for the sd-fp shell nuclei with effective charges $e_p$ = 1.5e and $e_n$ = 0.5e. Two different sets (SMI and SMII) of SM calculations with different truncations are performed to check the dependence of the calculated $b^2$ on the truncations. (${E_1}^{SM}$, ${E_2}^{SM}$), ($m_1$, $m_2$) and limits of $n_i$, $n_f$ and their values are different in the two sets. However, ${b_i}^2$, ${b_f}^2$ values obtained from SMI results (given ahead) are qualitatively similar to the ones obtained from SMII, only a slight quantitative difference is found (within limits of SMII results). 

 SMI calculations are performed with OXBASH code \cite{BAB} using sdpfmw Hamiltonian \cite{BAB} and SMII calculations are done with the NuShellX code \cite{BABN} using sdpfmwpn interaction \cite{BABN}. In both SMI and SMII calculations, for positive parity states, full sd valence space have been used. For negative parity states, 1p-1h excitations  
 [${(2s1d)}^{A-16-n}{(2p1f)}^{n}$], with n = 1 partitions \cite{ABS} are taken in SMII and almost same partitions are taken in SMI except for some particle restriction in $1d_{5/2}$ for $^{30}$P and $^{32}$S. Experimental data have been taken for E (T =0) from \cite{nndc}, and E(T=1) from \cite{nndc} \cite{MSA}. Experimental (E1) transition strengths ${B(E1)}_{expt}$ are obtained from the level lifetimes, branching ratios and multipole mixing ratios \cite{nndc,AB}. \\ 
The first observation was that, sets of (${b_i}^2$, ${b_f}^2$) obtained using Eq(15) and Eq(8) or Eq(19) with the sets of ($n_i$, $n_f$) values, namely, P(1.5, 1.0), S(7.915, 1.928), Cl(0.0, 2.0) and Ar(0.30, 0.0) and SM results for $m_1$,$m_2$ (Table I) predicted  M(E1) (central) values for the four self-conjugate nuclei, as -1.357, -1.629, -1.567 and +0.381, respectively, surprisingly close to the experimental values (Table I). Finer tuning, following the method discussed above, led to the results presented in Table I. \\

 It is revealing to compare the ${b_i}^2$, ${b_f}^2$ of Table I with the two limits, namely, ${b_i}^2$ with ${b_f}^2$ = 0 and ${b_f}^2$ with ${b_i}^2$ = 0 and $b_i$ = $b_f$ in Eq(20). Two terms of Eq(20) are competing, depending on the magnitudes and signs of $m_1$, $m_2$, particularly, when $m_1$ and $m_2$ have opposite signs. This may give rise to wide ranges for both ${b}^2$ s from very small to very large values. This situation can be eliminated using these limiting values of $b_i$, $b_f$.  For the nuclei (P, S, Cl, Ar) ${b_i}^2$ ($\%$) =  ${2.925}^{+1.686}_{-1.663}$ (minimum), ${164.728}^{+12.351}_{-11.904}$ ( maximum, all limiting values exceed 50$\%$, showing importance of contributions from both terms), ${0.394}^{+0.115}_{-0.114}$ (maximum), ${2.632}^{+0.198}_{-0.205}$ (maximum) for ${b_f}^2$ = 0 and ${b_f}^2$ ($\%$) = ${36.105}^{+20.805}_{-20.523}$ ( maximum, upper limit $>$ 50$\%$), ${2.253}^{+0.169}_{-0.163}$ (minimum), ${3.769}^{+1.061}_{-1.092}$ (maximum), ${6.891}^{+0.519}_{-0.536}$ (maximum) for ${b_i}^2$ = 0, respectively. Similarly, the limit $b_i$ = $b_f$ = $b$ gives, ${b}^2$ ($\%$) =  ${6.087}^{+3.926}_{-3.556}$, ${2.978}^{+0.231}_{-0.221}$, ${0.225}^{+0.064}_{-0.098}$, and ${1.016}^{+0.077}_{-0.080}$, respectively, for (P, S, Cl, Ar).\\

The (${b_i}^2$, ${b_f}^2$) in ($\%$)(central values only) and ($m_1$, $m_2$) $\times 10^{2}$from SMI are (4.128, 1.007)(0.99, -7.26), (1.0, 2.095)(10.29, 1.5), (0.410, 0.132) (0.807, -2.943), (0.398, 2.479)(1.47, 2.36), respectively, for (P, S, Cl and Ar) nuclei. With the $m_1$, $m_2$ values limiting values of (${b_i}^2$, ${b_f}^2$) can be calculated for SMI.\\ 
This shows that a reasonble SM calculation, with an isoscalar Hamiltonian, can predict semi-empirically the isospin mixing probability quite well if the prescribed method is followed. \\

Using Eq(20) and guided by the magnitude and sign of $m_1$, $m_2$, the sign of ${M(E1)}_{expt}$ can be selected (while fitting). One has to use the limiting values of (${b_i}^2$, ${b_f}^2$) given above. The signs obtained for SMII results are (-, -, -, +) for (P, S, Cl, Ar), respectively.\\

In Table II, the unperturbed energies $H_{11}$, $H_{22}$, the gap $\bar{\Delta}$ and $H_{12}$ for each doublet in the self-conjugate nuclei P, S, Cl and Ar are shown. One can see that the unperturbed gaps and mixing matrix elements are quite large. This is because of large values of ${\Delta}^{\prime}$ for each doublet. 
One can see from the table that except for $^{30}$P, isospin mixing matrix elements are larger for the lower doublets.\\ 

 For the second group of nuclei only one pair of mixed doublet is to be considered. Obviously an unique value of $b^2$ for this group can be obtained if experimental B(GT)/ forbidden B(E1) is known. In $^{54}Co$ \cite{Lise}, retarded M1 transition ${4_{2}}^{+} (predominantly, T = 0) \rightarrow 3^{+}$(T=0) occurs from the doublet ${4^+}_{1,2}$. 
We have considered three SM results for the energies of the mixed doublet, calculated in the fp valence space with different interactions and particle truncations. (${E_2}^{SM}$, ${E_1}^{SM}$) in keV are (2814, 2483) \cite{Lise}, (2934, 2683) (with fpd6pn interaction, \cite{BAB}), and (2839, 2562)(with fpd6npn interaction, \cite{BAB}) (present work). Experimental energies of the doublet are (2851.30, 2651.98) keV \cite{nndc}. The measured \cite{Lise} ${b_i}^2$ value = ${0.0023}^{+0.0029}_{-0.0010}$ can be reproduced exactly by the $n_i$ values (Eq(19)) within the limits $n_c\le n_i \le n_L$ for each SM results. Using Eq(6), Eq(7) and Eq(11), one can obtain  $H_{11} = {2652.44}^{+0.58}_{-0.20}$, $H_{22} = {2850.84}^{-0.58}_{+0.20}$, $\bar{\Delta}$ = ${198.40}^{-1.16}_{+0.40}$ and $H_{12}$ = ${9.55}^{+4.79}_{-2.37}$, all in keV.\\
  
Similarly, in  $^{64}Ge$ \cite{Fer},\cite{nndc}, the ${5_1}^- \rightarrow {4_1}^+$ forbidden E1 transition has been considered. 
Experimental and SM (with June45 interaction in NuShellX)  energies of the doublet are (2669.6, 2052.6) \cite{nndc} and (2343, 2115) keV, respectively. Measured ${b_f}^2$ values 0.012 (Ref.9 of \cite{Fer}) and 0.025 \cite{Fer} can be reproduced with $n_f$ = 1.42 and 1.36, respectively, which are within the range (-0.8630, 1.4724) for the SM result and are noted to be closer to the upper limit. However these measured values are not adopted \cite{nndc} yet.  
Thus a probable predicted value may be  around 0.0026 for $n_f$ = 1.46.

\begin{table*}%[h]
	\caption{\label{gt} Tabulation of predicted values of (${b_i}^2$,  ${b_f}^2$) for pairs of doublets in self-conjugate isotopes of P, S, Cl and  Ar. Relevant references for experimental energies and M(E1) values are discussed in the text. Theoretical energies  and ($m_1$,   $m_2$) values are from SMII calculations.} 

	\advance\leftskip-0cm
	\begin{tabular} {c| c|  c| c| c| c| c| c|}
		\hline
		\hline
Nucleus&J$_i^\pi$-J$_i^\pi$/&E$_1$, E$_2$ &E$_1^{SM}$, E$_2^{SM}$ &(n$_c \leq n \leq n_L$)& (m$_2$, m$_1$)&b$_i^2$/ b$_f^2$&Expt. M(E1)\\
Transition&J$_f^\pi$-J$_f^\pi$&Expt.&(keV)&&$\times 10^{-2}$&(\%)&$\times 10^{-2}$\\
&mixing&(keV)&&&&&\\

		\hline
		$^{30}$P&4$_1^-$- 4$_1^-$&(4231.97, 7057)&(4427, 6610)& (-8.66373, 2.54804)&(8.01, -2.28)&${4.723}^{+2.150}_{-2.243}$&$\pm{1.37}^{+0.35}_{-0.47}$\\
4$_1^-$- 3$_1^+$&3$_1^+$- 3$_1^+$&(1973.27, 5508.55)&(2062, 5657)&(0.25179, 15.15726)&&${2.462}^{+0.110}_{-0.092}$&\\
\hline
$^{32}$S&3$_1^-$- 3$_1^-$&(5006.2, 10221.2)&(5840, 9560)& (-21.55272, 8.66165)&(-1.27, 10.86)&${1.237}^{+0.210}_{-0.213}$&$\pm{1.63}^{+0.06}_{-0.06}$\\
3$_1^-$- 2$_1^+$&2$_1^+$- 2$_1^+$&(2230.57, 7115.30)&(2148, 7052)&(0.13210, 33.61898&&${2.688}^{+0.226}_{-0.221}$&\\
\hline
$^{34}$Cl&4$_1^-$- 4$_1^-$&(3600.27, 6207.1)&(3374, 5969)& (-5.58822, 0.02547)&(2.533, 0.819)&${0.237}^{+0.0530}_{-0.0545}$&$\pm{0.159}^{+0.021}_{-0.025}$\\
4$_1^-$- 3$_1^+$&3$_1^+$- 3$_1^+$&(146.36, 4717.4)&(133, 4774)&(1.61795, 107.33117)&&${0.192}^{+0.094}_{-0.092}$&\\
\hline
$^{36}$Ar&5$_1^-$- 5$_1^-$&(5171.13, 9014.9)&(4995, 8638)& (-6.58735, 0.36304)&(-2.33, -1.44)&${0.440}^{+0.067}_{-0.059}$&$\pm{0.378}^{+0.014}_{-0.015}$\\
5$_1^-$- 4$_1^+$&4$_1^+$- 4$_1^+$&(4414.40, 10460.6\footnote{see text})&(4564, 10311)&&&${2.460}^{+0.065}_{-0.098}$&\\
\hline
		\hline
		\hline
	\end{tabular}
	\end{table*}

The experimental GT matrix element for the transition from $^{24}Al^{m}$($1^+$,  T = 1) level at 369 keV to the 9828.11 keV,($1^+$, predominantly, T = 0) level of $^{24}Mg$ was  found \cite{Brown}
 to be much larger than the theory predicts. The authors of Ref. \cite{Brown} conjectured that this was very likely due to the isospin mixing with the closely lying 9967.19 keV, ($1^+$, predominantly, T=1) state which has a much larger GT matrix element. They \cite{Brown} have provided , $m_2$ = $<1^{+},T = 1|O(GT)|1^{+}, T = 1,Al>$ = 1.358 (effective) and = 1.732 (free) and 
$m_1$ = $<1^{+},T = 0|O(GT)|1^{+}, T = 1,Al>$ = 0.165 (effective) and = 0.215 (free) and ${M(GT)}_{expt}$ = 0.726 (0.103). We have  calculated ${b_f}^2$, using expression of the form $M(GT)_{expt}$ = $\sqrt{(1-{b_f}^2)}$ $m_1$ - $|b_f|$ $m_2$, for this doublet and found very large mixing. SM model calculation in the full sd valence space with w-interaction with OXBASH gives $\Delta = 3$ i.e. almost degenerate doublet. ${\Delta}^{\prime}$ = 139.08. So using $b^{2} = 0.5(1-{\Delta}/{\Delta}^{\prime})$ we get ${b_{f}}^{2}$ = 0.489215, very close to limiting value, ( ${b_{f}}^2$ =1/2 ).  
The measured and SM results for the energies of the doublet are (9828.11, 9967.19) \cite{nndc} and (9987, 9990) keV, respectively. The $n_f$ range is (-0.0165, 0.7489). The $n_f$ values ${0.1432}^{-0.1392}_{+0.1265}$ (eff) and ${0.3272}^{-0.0929}_{+0.0846}$ (free) give
${b_{f}}^2$  by fitting to the $M(GT)_{expt}$ as ${0.3957}^{+0.0909}_{-0.0827}$ (eff) and ${0.2755}^{+0.0607}_{-0.0552}$ (free), respectively. This amount of mixing is much larger than that (${20}_{-9.904}^{+9.142}$\%) found recently in a doublet in $ ^{26}$Si \cite{Liu}. However, ${H}_{12}$ = ${68.01}^{+1.50}_{-3.52}$   (eff) only \cite{Ray} since  ${\Delta}^{\prime}$ is small. With the obtained ${b_f}^2$ values, predictions for unknown M(GT) for the $^{24}Al^{m}$($1^+$,  T = 1) to $^{24}$Mg($1^+$, predominantly, T=1) transition, using ${M(GT)}_{expt}$ = $\sqrt{(1-{b_f}^2)}$ $m_2$ + $|b_f|$ $m_1$, are ${1.1595}^{-0.0714}_{+0.0584}$(eff) and ${1.5871}^{-0.0513}_{+0.0432}$ (free). $\bar{\Delta}$ = 3.73 for the upper limit of ${b_f}^2$ (eff) using Eq(7), which is very close to $\Delta$ = 3. Thus SM predicts ${b_f}^2$ very close to that obtained by fitting to the upper limit of ${M(GT)}_{expt}$.\\

In conclusion, it can be pointed out that a new  method for two-level model has been developed to calculate isospin mixing in nuclei using isoscalar Hamiltonian and minimum experimental inputs. The method is  general enough to be applicable to other two-level mixing problems. \\

\begin{table*}%[h]
	\caption{\label{gt} Unperturbed energies $H_{11}$, $H_{22}$, unperturbed gaps $\bar{\Delta}$ and isospin mixing matrix elements $H_{12}$ for initial (i) and final (f) set of doublets tabulated in keV are calculated with  ${b_i}^2$ and ${b_f}^2$  from Table I.} 
	\advance\leftskip-0cm
	\begin{tabular} {c| c|  c| c| c| c| }
		\hline
		\hline
Nucleus&J$_i^\pi$-J$_i^\pi$/&$ H_{11}^i$/ &$ H_{22}^i$/ &$\bar{\Delta}_i$/&H$_{12}^i$/\\
Transition&J$_f^\pi$-J$_f^\pi$&$H_{11}^f$ & $H_{22}^f$& $\bar{\Delta}_f$ & H$_{12}^f$ \\%\multispan{4}\hfil(keV)\hfil \\ %&(keV)&(keV)&(keV)\\
%&mixing&&&&\\
		\hline
		$^{30}$P&4$_1^-$- 4$_1^-$&${4365.40}^{+60.74}_{-63.37}$&${6923.57}^{-60.74}_{+63.37}$& ${2558.18}^{-121.48}_{+126.73}$&${599.27}^{+115.44}_{-159.94}$\\

4$_1^-$- 3$_1^+$&3$_1^+$- 3$_1^+$&${2060.31}^{+3.89}_{-3.25}$&${5421.51}^{-3.89}_{+3.25}$&${3361.20}^{-7.78}_{+6.50}$&${547.84}^{+11.79}_{+10.08}$\\
\hline
$^{32}$S&3$_1^-$- 3$_1^-$&${5070.71}^{+10.95}_{-11.11}$&${10156.69}^{-10.95}_{+11.11}$& ${5085.98}^{-21.90}_{22.22}$&${576.42}^{+46.35}_{-51.40}$\\

3$_1^-$- 2$_1^+$&2$_1^+$- 2$_1^+$&${2361.87}^{+11.04}_{-10.79}$&${6984.00}^{-11.04}_{+10.79}$&${4622.13}^{-22.08}_{+21.59}$&${790.02}^{+31.59}_{-32.31}$\\
\hline
$^{34}$Cl&4$_1^-$- 4$_1^-$&${3606.45}^{+1.38}_{-1.42}$&${6200.92}^{-1.38}_{+1.42}$& ${2594.47}^{-2.76}_{+2.84}$&${126.76}^{+13.42}_{-15.49}$\\

4$_1^-$- 3$_1^+$&3$_1^+$- 3$_1^+$&${155.14}^{+4.30}_{-4.20}$&${4708.62}^{-4.30}_{+4.20}$&${4553.49}^{-8.59}_{+8.41}$&${200.10}^{+44.00}_{-55.6210}$\\
\hline
$^{36}$Ar&5$_1^-$- 5$_1^-$&${5188.04}^{+2.57}_{-2.27}$&${8997.99}^{-2.57}_{+2.27}$& ${3809.94}^{-5.15}_{+4.54}$&${254.40}^{+18.59}_{-17.60}$\\

5$_1^-$- 4$_1^+$&4$_1^+$- 4$_1^+$&${4563.14}^{+3.93}_{-5.92}$&${10311.86}^{-3.93}_{+5.92}$&${5748.73}^{-7.86}_{+11.85}$&${936.57}^{+11.98}_{-18.38}$\\
\hline
		\hline
		\hline
	\end{tabular}
\end{table*}

%\begin {acknowledgment}

%\section*{Acknowledgment}
The Author sincerely acknowledges helps received from Professor M. Saha Sarkar for critical comments, computation and manuscript preparation. The Author also acknowledges Arkabrata Gupta and Dr. A. Bisoi for verification of a shell model calculation.
%\end {acknowledgment}


\begin{thebibliography}{50}
\bibitem{heisenberg} W. Heisenberg, Z. Phys. {\bf 77}, 1  (1932).
\bibitem{Bru} P. J. Brussaard and P. W. M. Glaudemans, Shell-Model Applications in Nuclear Spectroscopy, North-Holland Publishing Company, 1977.
\bibitem{Casten} R. F. Casten, Nuclear Structure from a Simple Perspective, General Editor P. E. Hodgson, Oxford University Press, 1990.
\bibitem {ekw} E. K. Warburton and J. Wesener, in: Isospin in Nuclear Physics, Editor D. H. Wilkinson, North-Holland Publishing Company, Amsterdam, 1969.
\bibitem {Lise} A. F. Lisetskiy, A. Schmidt, I. Schneider, C. Friessner, N. Pietralla, P.von Brentano, Phys. Rev. Lett. {\bf 89}, 012502 (2002).
\bibitem{BAB} B. A. Brown et al, MSU-NSCL Report No. 1289, 2004 (unpublished).
\bibitem {BABN} B. A. Brown, A version of NuShellX (private communication).
\bibitem{ABS} Abhijit Bisoi, Y. Sapkota, S. Sarkar, M. saha Sarkar, (presented in INPC-22, to be published in J. Phys. G : Conference Series, IOP publishing).
\bibitem{nndc} www.nndc.bnl.gov
\bibitem {MSA} M. S. Antony, A. Page, J. Britz, Atomic Data and Nuclear Tables, {\bf66}, 1-63 (1997).
\bibitem {AB} A. Bisoi, M.S. Sarkar, S. Sarkar, S. Ray, D. Pramanik, R. Kshetri, S. Nag, et al., Phys. Rev. C {\bf89}, 024303 (2014).
\bibitem {Fer} E. Farnea et al., Phys. Letts. B {\bf 551}, 56 (2003).
\bibitem {Brown} B. A. Brown and B. H. Wildenthal, Atomic Data and Nuclear Data Tables, {\bf33}, 347-404 (1985).
\bibitem{Liu} J. J. Liu et al, Phys. Rev. Lett. {\bf129}, 242502 (2022).
\bibitem {Ray} A. Ray, C. D. Hoyle and E. G. Adelberger, Nucl. Phys. A {\bf378}, 29 (1982).



\end{thebibliography}
\end{document}